\documentclass[aps,prl,twocolumn,showpacs,superscriptaddress,groupedaddress]{revtex4}  

\usepackage{graphicx}  
\usepackage{dcolumn}   
\usepackage{bm}        
\usepackage{amsmath}
\usepackage{amssymb}   
\usepackage{xcolor}    
\usepackage{latexsym}
\usepackage{amsfonts}

\definecolor{purple}{rgb}{0.5,0,0.5}
\definecolor{blue}{rgb}{0.0,0,0.9}
\definecolor{prdblue}{rgb}{0.133,0.118,0.498}
\usepackage[colorlinks=true, pdfstartview=FitV, linkcolor=prdblue, citecolor= prdblue, urlcolor=prdblue]{hyperref}

\begin{document}

\title{Multidimensional, high precision measurements of beam single spin asymmetries  in semi-inclusive $\pi^{+}$  electroproduction off protons in the valence region}

\newcommand*{\ANL}{Argonne National Laboratory, Argonne, Illinois 60439}
\newcommand*{\ANLindex}{1}
\affiliation{\ANL}
\newcommand*{\ASU}{Arizona State University, Tempe, AZ 85281}
\newcommand*{\ASUindex}{2}
\affiliation{\ASU}
\newcommand*{\SACLAY}{IRFU, CEA, Universit\'{e} Paris-Saclay, F-91191 Gif-sur-Yvette, France}
\newcommand*{\SACLAYindex}{3}
\affiliation{\SACLAY}
\newcommand*{\CNU}{Christopher Newport University, Newport News, Virginia 23606}
\newcommand*{\CNUindex}{4}
\affiliation{\CNU}
\newcommand*{\UCR}{University of California, Riverside, CA 92521}
\newcommand*{\UCRindex}{5}
\affiliation{\UCR}
\newcommand*{\UCONN}{University of Connecticut, Storrs, Connecticut 06269}
\newcommand*{\UCONNindex}{6}
\affiliation{\UCONN}
\newcommand*{\DUKE}{Duke University, Durham, North Carolina 27708-0305}
\newcommand*{\DUKEindex}{7}
\affiliation{\DUKE}
\newcommand*{\DUQUESNE}{Duquesne University, 600 Forbes Avenue, Pittsburgh, PA 15282 }
\newcommand*{\DUQUESNEindex}{8}
\affiliation{\DUQUESNE}
\newcommand*{\FU}{Fairfield University, Fairfield CT 06824}
\newcommand*{\FUindex}{9}
\affiliation{\FU}
\newcommand*{\FERRARAU}{Universita' di Ferrara , 44121 Ferrara, Italy}
\newcommand*{\FERRARAUindex}{10}
\affiliation{\FERRARAU}
\newcommand*{\FIU}{Florida International University, Miami, Florida 33199}
\newcommand*{\FIUindex}{11}
\affiliation{\FIU}
\newcommand*{\FSU}{Florida State University, Tallahassee, Florida 32306}
\newcommand*{\FSUindex}{12}
\affiliation{\FSU}
\newcommand*{\GWUI}{The George Washington University, Washington, DC 20052}
\newcommand*{\GWUIindex}{13}
\affiliation{\GWUI}
\newcommand*{\ISU}{Idaho State University, Pocatello, ID 83209}
\newcommand*{\ISUindex}{14}
\affiliation{\ISU}
\newcommand*{\INFNFE}{INFN, Sezione di Ferrara, 44100 Ferrara, Italy}
\newcommand*{\INFNFEindex}{15}
\affiliation{\INFNFE}
\newcommand*{\INFNFR}{INFN, Laboratori Nazionali di Frascati, 00044 Frascati, Italy}
\newcommand*{\INFNFRindex}{16}
\affiliation{\INFNFR}
\newcommand*{\INFNGE}{INFN, Sezione di Genova, 16146 Genova, Italy}
\newcommand*{\INFNGEindex}{17}
\affiliation{\INFNGE}
\newcommand*{\INFNRO}{INFN, Sezione di Roma Tor Vergata, 00133 Rome, Italy}
\newcommand*{\INFNROindex}{18}
\affiliation{\INFNRO}
\newcommand*{\INFNTUR}{INFN, Sezione di Torino, 10125 Torino, Italy}
\newcommand*{\INFNTURindex}{19}
\affiliation{\INFNTUR}
\newcommand*{\INFNPAV}{INFN, Sezione di Pavia, 27100 Pavia, Italy}
\newcommand*{\INFNPAVindex}{20}
\affiliation{\INFNPAV}
\newcommand*{\ORSAY}{Universit'{e} Paris-Saclay, CNRS/IN2P3, IJCLab, 91405 Orsay, France}
\newcommand*{\ORSAYindex}{21}
\affiliation{\ORSAY}
\newcommand*{\Juelich}{Institute fur Kernphysik (Juelich), Juelich, Germany}
\newcommand*{\Juelichindex}{22}
\affiliation{\Juelich}
\newcommand*{\JMU}{James Madison University, Harrisonburg, Virginia 22807}
\newcommand*{\JMUindex}{23}
\affiliation{\JMU}
\newcommand*{\KNU}{Kyungpook National University, Daegu 41566, Republic of Korea}
\newcommand*{\KNUindex}{24}
\affiliation{\KNU}
\newcommand*{\LAMAR}{Lamar University, 4400 MLK Blvd, PO Box 10046, Beaumont, Texas 77710}
\newcommand*{\LAMARindex}{25}
\affiliation{\LAMAR}
\newcommand*{\MIT}{Massachusetts Institute of Technology, Cambridge, Massachusetts  02139-4307}
\newcommand*{\MITindex}{26}
\affiliation{\MIT}
\newcommand*{\MISS}{Mississippi State University, Mississippi State, MS 39762-5167}
\newcommand*{\MISSindex}{27}
\affiliation{\MISS}
\newcommand*{\ITEP}{National Research Centre Kurchatov Institute - ITEP, Moscow, 117259, Russia}
\newcommand*{\ITEPindex}{28}
\affiliation{\ITEP}
\newcommand*{\UNH}{University of New Hampshire, Durham, New Hampshire 03824-3568}
\newcommand*{\UNHindex}{29}
\affiliation{\UNH}
\newcommand*{\NMSU}{New Mexico State University, PO Box 30001, Las Cruces, NM 88003, USA}
\newcommand*{\NMSUindex}{30}
\affiliation{\NMSU}
\newcommand*{\NSU}{Norfolk State University, Norfolk, Virginia 23504}
\newcommand*{\NSUindex}{31}
\affiliation{\NSU}
\newcommand*{\OHIOU}{Ohio University, Athens, Ohio  45701}
\newcommand*{\OHIOUindex}{32}
\affiliation{\OHIOU}
\newcommand*{\ODU}{Old Dominion University, Norfolk, Virginia 23529}
\newcommand*{\ODUindex}{33}
\affiliation{\ODU}
\newcommand*{\PEN}{Science Division, Penn State University Berks, Reading, Pennsylvania 19610, USA}
\newcommand*{\PENindex}{34}
\affiliation{\PEN}
\newcommand*{\JLUGi}{II. Physikalisches Institut der Universit\"at Gie{\ss}en, 35392 Gie{\ss}en, Germany}
\newcommand*{\JLUGiindex}{35}
\affiliation{\JLUGi}
\newcommand*{\RPI}{Rensselaer Polytechnic Institute, Troy, New York 12180-3590}
\newcommand*{\RPIindex}{36}
\affiliation{\RPI}
\newcommand*{\URICH}{University of Richmond, Richmond, Virginia 23173}
\newcommand*{\URICHindex}{37}
\affiliation{\URICH}
\newcommand*{\ROMAII}{Universita' di Roma Tor Vergata, 00133 Rome Italy}
\newcommand*{\ROMAIIindex}{38}
\affiliation{\ROMAII}
\newcommand*{\SPS}{School of Physics, Southeast University, Nanjing 211189, Jiangsu, China}
\newcommand*{\SPSindex}{39}
\affiliation{\SPS}
\newcommand*{\SPI}{School of Physics and Institute for Nonperturbative Physics, Nanjing University, Nanjing 210093, Jiangsu, China}
\newcommand*{\SPIindex}{40}
\affiliation{\SPI}
\newcommand*{\SSN}{School of Science, Nanjing University of Posts and Telecommunications, Nanjing 210023, Jiangsu, China}
\newcommand*{\SSNindex}{41}
\affiliation{\SSN}
\newcommand*{\MSU}{Skobeltsyn Institute of Nuclear Physics, Lomonosov Moscow State University, 119234 Moscow, Russia}
\newcommand*{\MSUindex}{42}
\affiliation{\MSU}
\newcommand*{\SCAROLINA}{University of South Carolina, Columbia, South Carolina 29208}
\newcommand*{\SCAROLINAindex}{43}
\affiliation{\SCAROLINA}
\newcommand*{\TEMPLE}{Temple University,  Philadelphia, PA 19122 }
\newcommand*{\TEMPLEindex}{44}
\affiliation{\TEMPLE}
\newcommand*{\JLAB}{Thomas Jefferson National Accelerator Facility, Newport News, Virginia 23606}
\newcommand*{\JLABindex}{45}
\affiliation{\JLAB}
\newcommand*{\UTFSM}{Universidad T\'{e}cnica Federico Santa Mar\'{i}a, Casilla 110-V Valpara\'{i}so, Chile}
\newcommand*{\UTFSMindex}{46}
\affiliation{\UTFSM}
\newcommand*{\INSUBRIA}{Universit\`{a} degli Studi dell'Insubria, 22100 Como, Italy}
\newcommand*{\INSUBRIAindex}{47}
\affiliation{\INSUBRIA}
\newcommand*{\BRESCIA}{Universit\`{a} degli Studi di Brescia, 25123 Brescia, Italy}
\newcommand*{\BRESCIAindex}{48}
\affiliation{\BRESCIA}
\newcommand*{\GLASGOW}{University of Glasgow, Glasgow G12 8QQ, United Kingdom}
\newcommand*{\GLASGOWindex}{49}
\affiliation{\GLASGOW}
\newcommand*{\YORK}{University of York, York YO10 5DD, United Kingdom}
\newcommand*{\YORKindex}{50}
\affiliation{\YORK}
\newcommand*{\VIRGINIA}{University of Virginia, Charlottesville, Virginia 22901}
\newcommand*{\VIRGINIAindex}{51}
\affiliation{\VIRGINIA}
\newcommand*{\WM}{College of William and Mary, Williamsburg, Virginia 23187-8795}
\newcommand*{\WMindex}{52}
\affiliation{\WM}
\newcommand*{\YEREVAN}{Yerevan Physics Institute, 375036 Yerevan, Armenia}
\newcommand*{\YEREVANindex}{53}
\affiliation{\YEREVAN}

\newcommand*{\NOWBRESCIA}{Universit\`{a} degli Studi di Brescia, 25123 Brescia, Italy}

\author {S.~Diehl} 
\affiliation{\JLUGi}
\affiliation{\UCONN}
\author {A.~Kim} 
\affiliation{\UCONN}
\author {G.~Angelini} 
\affiliation{\GWUI}
\author {K.~Joo} 
\affiliation{\UCONN}
\author {S.~Adhikari} 
\affiliation{\FIU}
\author {M.~Amaryan} 
\affiliation{\ODU}
\author {M.~Arratia} 
\affiliation{\UCR}
\author {H.~Atac} 
\affiliation{\TEMPLE}
\author {H.~Avakian} 
\affiliation{\JLAB}
\author {C.~Ayerbe Gayoso} 
\affiliation{\WM}
% ...


\collaboration{The CLAS Collaboration}
\noaffiliation

\begin{abstract}
High precision measurements of the polarized electron beam-spin asymmetry in semi-inclusive deep inelastic scattering (SIDIS) from the proton have been performed using a 10.6~GeV incident electron beam and the CLAS12 spectrometer at Jefferson Lab. We report here a high precision multidimensional study of single $\pi^{+}$ SIDIS data over a large kinematic range in Bjorken x, fractional energy and transverse momentum of the hadron as well as photon virtualities $Q^{2}$ ranging from $1-7\,$GeV$^{2}$. In particular, the structure function ratio $F^{\sin\phi}_{LU}/F_{UU}$ has been determined, where $F^{\sin\phi}_{LU}$ is a twist-3 quantity that can reveal novel aspects of emergent hadron mass and quark-gluon correlations within the nucleon. The data's impact on the evolving understanding of the underlying reaction mechanisms and their kinematic variation is explored using theoretical models for the different contributing transverse momentum dependent parton distribution functions.
\end{abstract}

\maketitle


Since the discovery of quarks more than fifty years ago \cite{Bloom:1969kc, Breidenbach:1969kd}, science has sought to understand what they are and how they are bound by gluons to form nucleons.  Ensuing decades of measurements via deep inelastic scattering (DIS) of lepton beams off nucleons have charted the nucleon's gluon and quark momentum distributions in terms of one-dimensional (1-D) parton distribution functions (PDFs) \cite{GHR18, EN20, KNS20}.  These measurements have led to significant insights; but many structural aspects are invisible in such 1-D images, which are essentially obtained by averaging over all degrees-of-freedom except longitudinal momentum.  For instance, they cannot fully reveal whether quarks undergo orbital motion; if there is a connection between quark motion, their spin and the spin of the proton; or how the proton's total spin is built from the orbital angular momenta and spins of the gluons and quarks within.  Today, building upon twenty-five years of theory \cite{AKo95, MT96, BM98, GMS05, Bacch07, Metz:2016swz}, new possibilities for 3-D imaging exist, which promise to deliver answers to these and other basic questions.  In the past decade, a mathematical framework has been developed that links information on the confined motion of partons inside a rapidly moving nucleon with transverse momentum dependent parton distribution functions (TMDs) \cite{Bacch07, GMS05, BM98, PR19}.  Knowledge of TMDs reveals, \emph{inter alia}: the orbital motion of quarks within the parent nucleon; correlations between parton momentum and spin; and additional effects driven by emergent hadron mass \cite{Roberts:2021xnz, Roberts:2021nhw}, e.g., dynamical chiral symmetry breaking (DCSB) \cite{Accardi:2020iqn} and the generation of quark+quark (diquark) correlations within nucleons \cite{Barabanov:2020jvn}.

\begin{figure}[b!]
	\centering
		\includegraphics[width=0.40\textwidth]{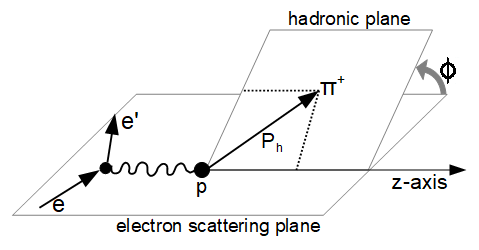}
	\caption{Schematic diagram of the reaction kinematics of the single pion semi-inclusive deep inelastic scattering process.}
	\label{fig:SIDIS_process}
\end{figure}

A powerful tool for studying nucleon structural distributions in the plane transverse to its light-front (longitudinal) direction of motion is provided by semi-inclusive DIS (SIDIS), which involves measuring at least one specified hadron in the final state plus the scattered lepton, see Fig.\,\ref{fig:SIDIS_process}.
Spin asymmetries in polarized SIDIS are related to convolutions of TMDs with fragmentation functions (FFs), the latter parametrizing the probability for a given struck parton to emit a particular hadron \cite{Metz:2016swz}.
Such asymmetries attract intense interest \cite{DWS91, EGS03, Bacch04, MS04, Anselmino:2013vqa, Avakian:2019drf, Review20}.

Sizable single spin asymmetries (SSAs) have been observed in SIDIS with longitudinally polarized lepton beams and unpolarized targets (hereafter described as beam SSAs) \cite{Aira00, HERMES:2001hbj, Aira07, Adol14, Gohn14, Aira19}.
Beam SSAs are twist-3 objects, \emph{i.e}.\ compared with many other processes, their cross-sections are suppressed by ${\cal O}(M/Q)$, where $M$ is the target nucleon mass and $Q^2$ is the photon virtuality.  This property can make measuring them difficult.
However, at the energies characteristic of existing fixed-target (as opposed to colliding beam) facilities, contributions of ${\cal O}(M/Q)$ can be significant, making beam SSAs accessible.
One thereby gains rare access to information  about: correlations between gluons and quarks within the target
%
\cite{Avakian:2019drf, Review20}; and emergent hadron mass, which is responsible for the bulk of observable mass in the Universe \cite{Roberts:2021xnz, Roberts:2021nhw, Accardi:2020iqn}.

This Letter presents high-precision beam SSAs measured
in $\pi^+$ SIDIS of longitudinally polarized electrons off unpolarized protons with a wide range of fully differential multidimensional kinematics 
on $Q^2\in [1.7,7.0]\,$GeV$^2$, 
$x_B\in [0.13,0.52]$,
$z\in[0.17,0.7]$, 
and $P_{T}$ up to 0.85\,GeV (c = 1).
Here,
$Q^2$ is the momentum transferred into the system by the lepton probe (the photon virtuality);
$x_B$ is the fraction of the proton's momentum carried by the struck quark;
$P_{T}$ is the hadron's transverse momentum, with respect to the virtual photon;
$y$ is the energy fraction of the incoming lepton carried by the virtual photon and $z$ is the fraction of the virtual photon's energy carried by the outgoing hadron in the lab frame.
The reaction kinematics of the process are sketched in Fig.\,\ref{fig:SIDIS_process}.

In the one-photon exchange approximation beam SSAs ($A_{LU}$) are defined thus:
\begin{equation}
\label{eq:BSA}
\begin{split}
	A_{LU}(z, P_T, \phi, & x_{B}, Q^{2})  = \frac{d\sigma^{+} - d\sigma^{-}}{d\sigma^{+} + d\sigma^{-}}  \\
	&= \frac{A_{LU}^{\sin\phi} \sin\phi}{1 + A_{UU}^{\cos\phi} \cos\phi + A_{UU}^{\cos2\phi} \cos2\phi},
\end{split}
\end{equation}
where $d\sigma^{\pm}$ is the differential cross section for each beam helicity state ($\pm$): spin parallel/antiparallel to the beam direction.
%
%
The subscripts of the moments $A_{ij}$ represent the longitudinally polarized (L) or unpolarized (U) state of the beam and target, respectively. 
%
$\phi$ is the azimuthal angle between the electron scattering plane and the hadronic reaction plane, see Fig.\,\ref{fig:SIDIS_process}.

Our chief focus is the $\sin\phi$ moment, $A_{LU}^{\sin\phi}$, which provides access to dynamical aspects of proton structure, 
as will become clear.  It is proportional to the polarized structure function $F_{LU}^{\sin\phi}$:
\begin{equation}\label{eq:ALU}
	A_{LU}^{\sin\phi} = \frac{\sqrt{2 \epsilon (1 - \epsilon)}~F_{LU}^{\sin\phi}}{F_{UU,T} + \epsilon F_{UU,L}},
\end{equation}
where the terms in $F_{UU} = F_{UU,T} + \epsilon F_{UU,L}$ are the contributions from longitudinal and transverse polarizations of the virtual photon, and $\epsilon$ is the ratio of their fluxes.

%
A TMD interpretation of our data requires that a (factorized) convolution formula be a valid interpretative tool \cite{Bacch07, Leve94}.  The required kinematic conditions might not be met at our largest $P_T$ values in the smallest $x_B$ bins and a complementary analysis framework may be applicable on this domain \cite{Bacchetta:2008xw}; but contemporary theory cannot provide rigorous guidance on these points because factorization has not yet been proved in connection with twist-3 observables \cite{Bacchetta:2019qkv}, although progress in that direction is being made \cite{Feige:2017zci, Balitsky:2017flc, Moult:2019mog}.  Consequently, we proceed by assuming factorization is valid and remain vigilant against manifest violations.  So, we write \cite{Bacch07, Leve94}:
\begin{eqnarray}
\label{eqn:FLU_full}
	F_{LU}^{\sin\phi} = \frac{2M}{Q}\;  {\cal C} \Bigg[ -\frac{\hat{\textbf{h}} \cdot \textbf{k}_{T}}{M_{h}} \left(x_B e H^{\bot}_{1} + \frac{M_{h}}{M} f_{1} \frac{\tilde{G}^{\bot}}{z}\right) \nonumber \\
	+\frac{\hat{\textbf{h}} \cdot \textbf{p}_{T}}{M} \left(x_B g^{\bot} D_{1} + \frac{M_{h}}{M} h^{\bot}_{1} \frac{\tilde{E}}{z} \right) \Bigg],
\end{eqnarray}
where $\cal C$ indicates a convolution of TMDs and FFs.
Here $e$ is a twist-3 TMD, $H^{\bot}_{1}$ is the Collins FF,
$f_{1}$ is the unpolarized distribution function, $\tilde{G}^{\bot}$ is a twist-3 FF,
$g^{\bot}$ is a twist-3 T-odd distribution function, $D_{1}$ is the unpolarized FF,
$h^{\bot}_{1}$ is the Boer-Mulders function and $\tilde{E}$ is a twist-3 FF.
(The properties of these functions are detailed elsewhere \cite{Metz:2016swz, Avakian:2019drf, Review20}.)
Furthermore, $\textbf{p}_{T}$  ($\textbf{k}_{T}$) is the intrinsic quark transverse momentum in the generic distribution function $f_{1}$ (fragmentation function $D_{1}$), $M_{h}$ is the pion mass and $\hat{\textbf{h}}$ is a unit vector in the direction of the pion's transverse momentum.

Notably, most twist-3 structure functions can be separated into three terms using QCD's equations of motion: a twist-2 piece, relating to some single-parton density; a genuine twist-3 term, containing information on quark-gluon correlations and DCSB; and a term proportional to the current-quark mass, which is usually neglected for light quarks.  The so-called Wandzura-Wilczek (WW) approximation keeps only the twist-two piece \cite{Wandzura:1977qf}.  Crucially, the structure function $F_{LU}^{\sin\phi}$ is special because it contains no such twist-two contribution, \emph{i.e}.\ it is genuinely twist-three \cite{Bacch07}; hence, particularly sensitive to quark-gluon correlations.  Any analysis of our experiment that uses the WW approximation will return zero for the BSA \cite{Bastami:2018xqd}, in clear conflict with the data.

Since the several percent magnitude of the observed asymmetry cannot be explained by perturbative QCD, several nonperturbative mechanisms have been proposed.
One involves the $e H_1^\perp$ term~\cite{Efremov:2002qh, Cebulla:2007ej}, attributing the asymmetry to a coupling between the Collins FF $H_1^\perp$ and $e(x)$, which is a chiral-odd TMD; hence, sensitive to DCSB \cite{Roberts:2021xnz, Roberts:2021nhw, Accardi:2020iqn}.
Other mechanisms involve convolution of the Boer-Mulders function $h_1^\perp$ with the FF $\tilde{E}$ and the coupling between the unpolarized distribution function $f_{1}$ and the twist-3 FF  $\tilde{G}^{\bot}$.
Apart from those mentioned above, a mechanism involving the poorly known twist-3 TMD $g^\perp$ can also generate the beam SSA. $g^\perp$ appears in the decomposition of the quark correlator if the dependence on the light-cone vector is included and is sensitive to target quark-gluon correlations.
To model the twist-3 T-odd chiral-even TMD $g^\perp$ it is necessary to include final state interactions, which can be estimated via one-gluon exchange.
Therefore, studying beam SSAs provides a unique opportunity to unravel the role of genuine twist-3 effects; and the subsequent discussion suggests that our data are particularly sensitive to the $e H^{\bot}_{1}$ (DCSB) and $g^{\bot} D_{1}$ (quark gluon correlation) terms in Eq.\,\eqref{eqn:FLU_full}.

\begin{figure}[t!]
	\centering
		\includegraphics[width=0.48\textwidth]{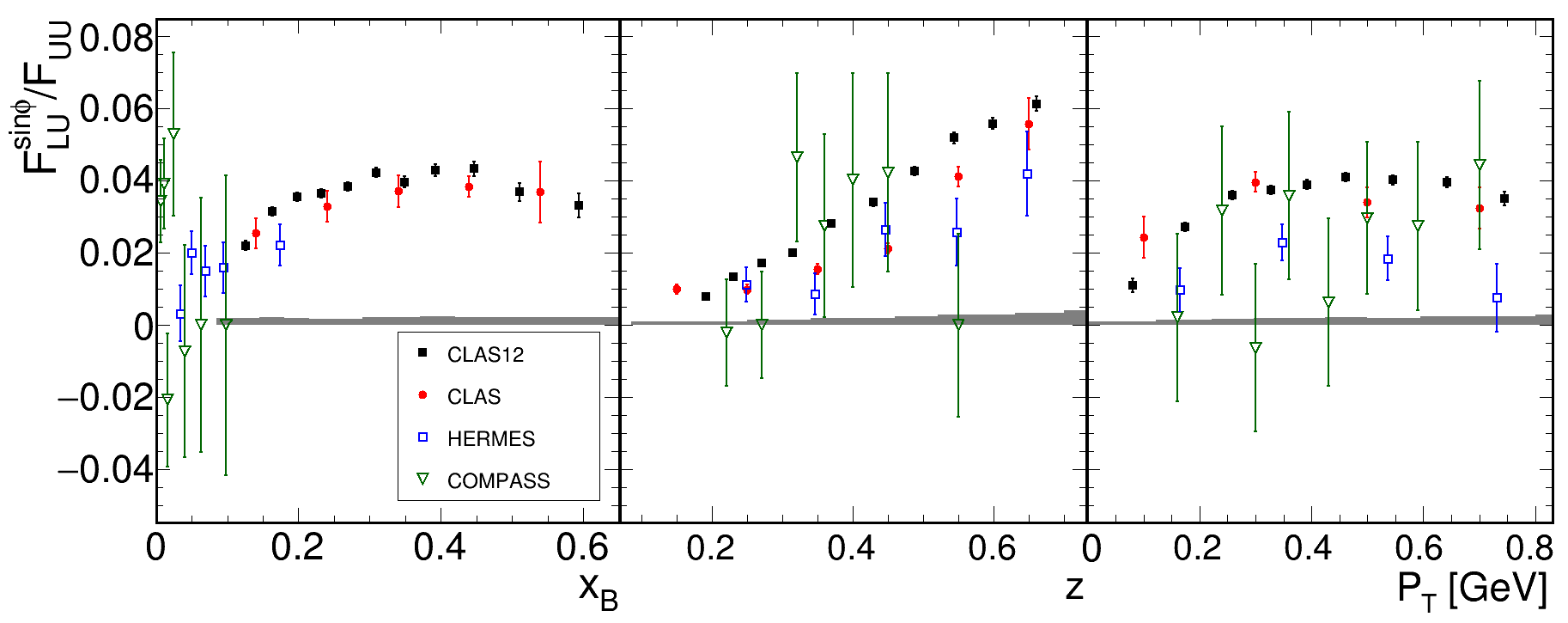}
	\caption{Status of current world data on $F^{\sin\phi}_{LU}/F_{UU}$ for $\pi^{+}$ in terms of kinematics and accuracy from CLAS12 (filled black squares), HERMES \cite{Aira19} (open blue squares), COMPASS \cite{Adol14} (all positive hadrons considered, open green triangles) and CLAS \cite{Gohn14} (filled red circles) as a function of $x_{B}$,  $z$ and $P_{T}$ integrated over all other kinematic variables. The $A^{\sin\phi}_{LU}$ values stated in references \cite{Adol14} and \cite{Gohn14} were transformed to $F^{\sin\phi}_{LU}/F_{UU}$ following Eq. (\ref{eq:ALU}). The grey histogram shows the systematic uncertainty of the present CLAS12 data. }
	\label{fig:comparison_1d}
\end{figure}

SIDIS $\pi^+$ electroproduction was measured at Jefferson Lab with CLAS12 (CEBAF Large Acceptance Spectrometer for experiments at 12 GeV) \cite{VDB20}. Beam SSAs were extracted over a wide range in $Q^{2}$, $x_{B}$, $z$, $P_{T}$ and $\phi$. 
The incident $10.6$\,GeV electron beam was longitudinally polarized and the target was unpolarized liquid hydrogen. The CLAS12 forward detector consists of six identical sectors within a toroidal magnetic field. The momentum and the charge of the particles were determined by 3 regions of drift chambers from the curvature of the particle trajectories in the magnetic field. The electron identification was based on a lead-scintillator electromagnetic sampling calorimeter in combination with a Cherenkov counter. Positive pions were identified by time-of-flight measurements.
For the selection of deeply inelastic scattered electrons, cuts on $Q^{2} > 1\, {\rm GeV}^{2}$, $y < 0.75$ and on the invariant mass of the hadronic final state $W > 2$~GeV, were applied. In addition, it was required that the $e'\pi^{+}X$ missing mass be larger than 1.5\,GeV to reduce the contribution from exclusive channels.

Fig.\,\ref{fig:comparison_1d} shows the new CLAS12 data as a function of $x_{B}$,  $z$, $P_{T}$, integrated over all other kinematic variables, compared with available world data for $F_{LU}^{\sin\phi}/F_{UU}$ from previous experiments. Details on the CLAS12 multidimensional analysis follow.  
Although $F_{LU}^{\sin\phi}$ was studied at HERMES \cite{Aira07,Aira19}, COMPASS \cite{Adol14} and CLAS \cite{Avak04, Gohn14} during the last two decades, there is still no consistent understanding of the contribution from each part to the total structure function. The high statistics on an extended kinematic range, which distinguishes the new data, enables a high precision multidimensional analysis; hence, provides an excellent basis for TMD and FF extraction.

\begin{figure}[t!]
	\centering
		\includegraphics[width=0.240\textwidth]{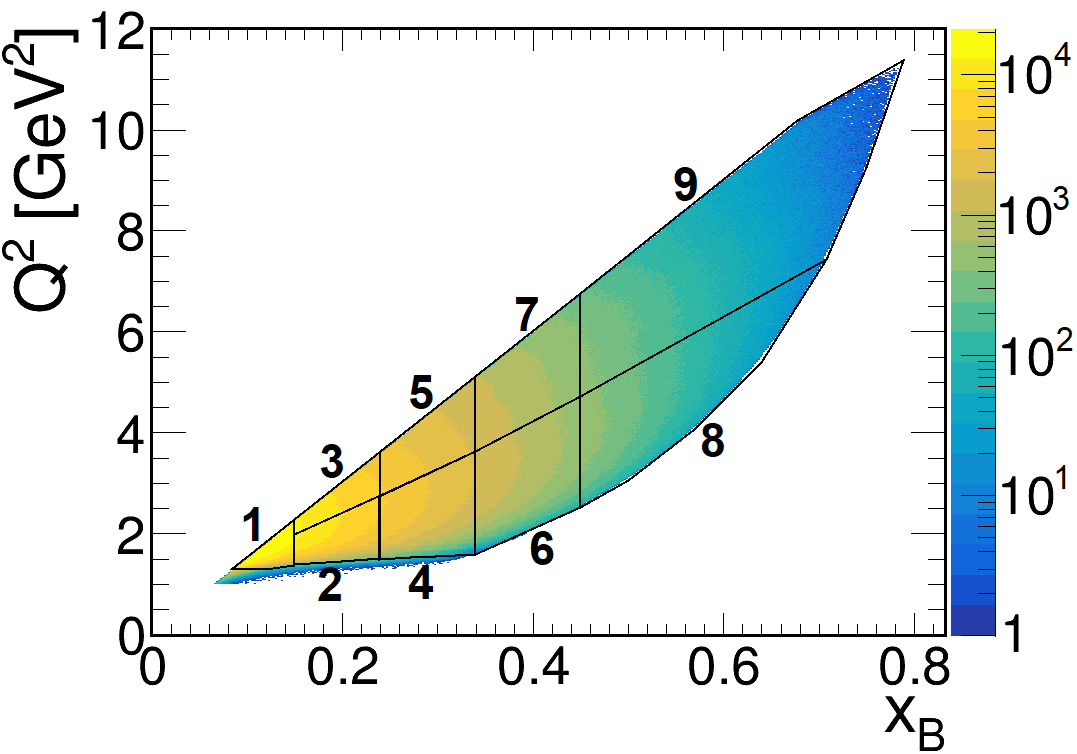}
		\includegraphics[width=0.233\textwidth]{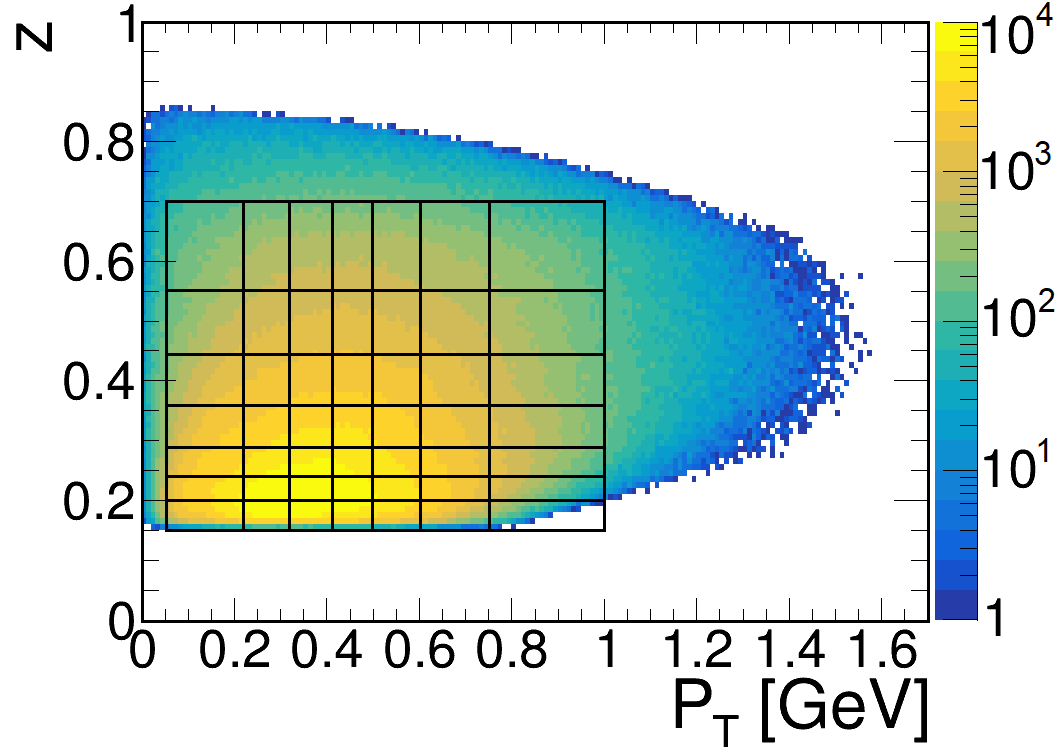}
	\caption{Left: Distribution of $Q^{2}$ versus $x_{B}$ with bin numbering and boundaries. Right: Correlation between $z$ and $P_{T}$ for $Q^{2}$ - $x_{B}$ - bin 1. The black lines indicate the bin borders. 
	\label{fig:binning}}
\end{figure}

For the multidimensional binning, first the electron variables are sorted in 9 bins in $Q^{2}$ and $x_{B}$ (see Fig.\,\ref{fig:binning}). For each of these $Q^{2}$ - $x_{B}$ bins, a binning is applied to $z$ and $P_{T}$ as exemplified in Fig.\,\ref{fig:binning}.

The beam SSA and its statistical uncertainty were determined experimentally from the number of counts with positive and negative helicity ($N^{\pm}_{i}$) in a specific bin $i$ as:
\begin{eqnarray}
	A_{LU} = \frac{1}{P_{e}} \frac{N^{+}_{i} - N^{-}_{i}}{N^{+}_{i} + N^{-}_{i}}~,
	~\sigma_{A_{LU}} = \frac{1}{P_{e}}\sqrt{\frac{1-(P_{e}~A_{LU})^{2}}{N^{+}_{i}+N^{-}_{i}}},~
\end{eqnarray}
\newline where $P_{e}$ is the average magnitude of the beam polarization. $P_{e}$ was measured with a M{\o}ller polarimeter upstream of CLAS12 and was 86.3\%$\pm$2.6\%. The polarization was flipped at 30~Hz
to minimize systematic effects.

To extract $A_{LU}^{\sin \phi}$, the beam SSA was measured as a function of the azimuthal angle $\phi$. Then the data was fit with a $\sin\phi$ function. The obtained $A_{LU}^{\sin \phi}$ moment is then related to $F^{\sin\phi}_{LU}/F_{UU}$ via Eq.~\eqref{eq:ALU}.
Several sources of systematic uncertainty were investigated, including beam polarization, radiative effects, particle identification and contamination from baryon resonances and exclusive $\rho$ meson production. A detailed Monte Carlo simulation was performed to study acceptance and bin-migration effects, which were both found to be negligible compared to the other contributions. The influence of additional azimuthal modulations $\cos\phi$ and $\cos2\phi$ on the extracted $\sin\phi$ amplitude was also evaluated, and found to be negligible.
The total point-to-point systematic uncertainty of $F^{\sin\phi}_{LU}/F_{UU}$, defined as the square-root of the quadratic sum of the uncertainties from all sources, is typically on the order of 5.6\% and dominated by the uncertainty from radiative effects (3.0\%) and acceptance and bin migration effects (2.7\%). The beam polarization adds an additional 3.0\% scale uncertainty to our observable.

\begin{figure}[t!]
	\centering
		\includegraphics[width=0.48\textwidth]{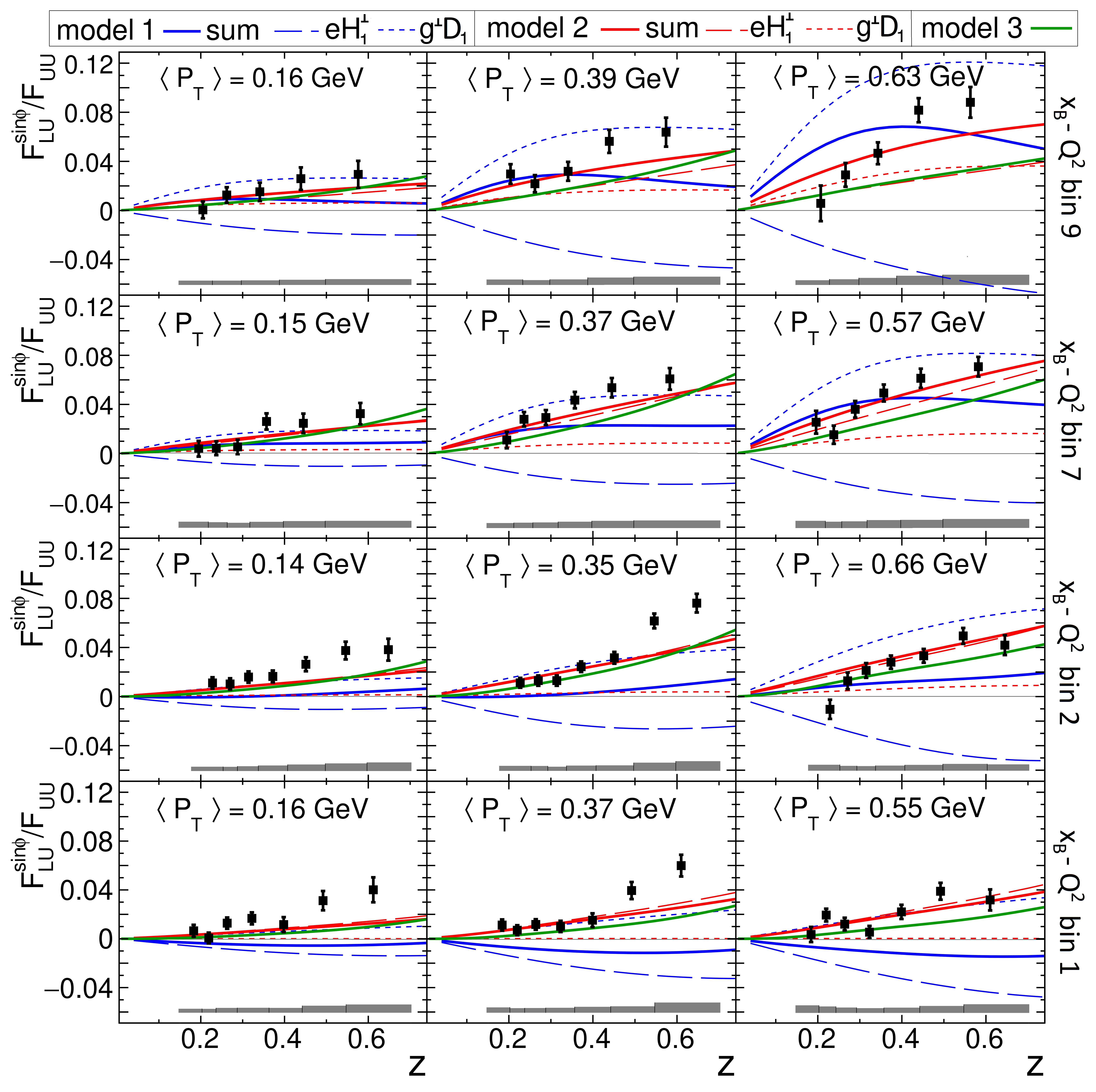}
	\caption{$z$ dependence of $F^{\sin\phi}_{LU}/F_{UU}$ for increasing $P_{T}$ bins (left to right) and for different $Q^{2}$-$x_{B}$ bins (bin 1: $\left\langle Q^{2}\right\rangle$ = 1.71 GeV$^{2}$, $\left\langle x_{B}\right\rangle$ = 0.13, bin 2: $\left\langle Q^{2}\right\rangle$ = 2.02 GeV$^{2}$, $\left\langle x_{B}\right\rangle$ = 0.19, bin 7: $\left\langle Q^{2}\right\rangle$ = 4.89 GeV$^{2}$, $\left\langle x_{B}\right\rangle$ = 0.39, bin 9: $\left\langle Q^{2}\right\rangle$ = 6.55 GeV$^{2}$, $\left\langle x_{B}\right\rangle$ = 0.52). The systematic uncertainty is given by the grey histogram. The predictions of the different theoretical models are shown by the bold lines (blue: model 1, red: model 2, magenta: model 3). For models 1 and 2 the contribution from $e H^{\bot}_{1}$ (long dashed line) and $g^{\bot} D_{1}$ (short dashed line) are shown in the same color as the final result.
\label{fig:FLU_z_dep_multidim}}
	
\end{figure}

\begin{figure}[t!]
	\centering
		\includegraphics[width=0.48\textwidth]{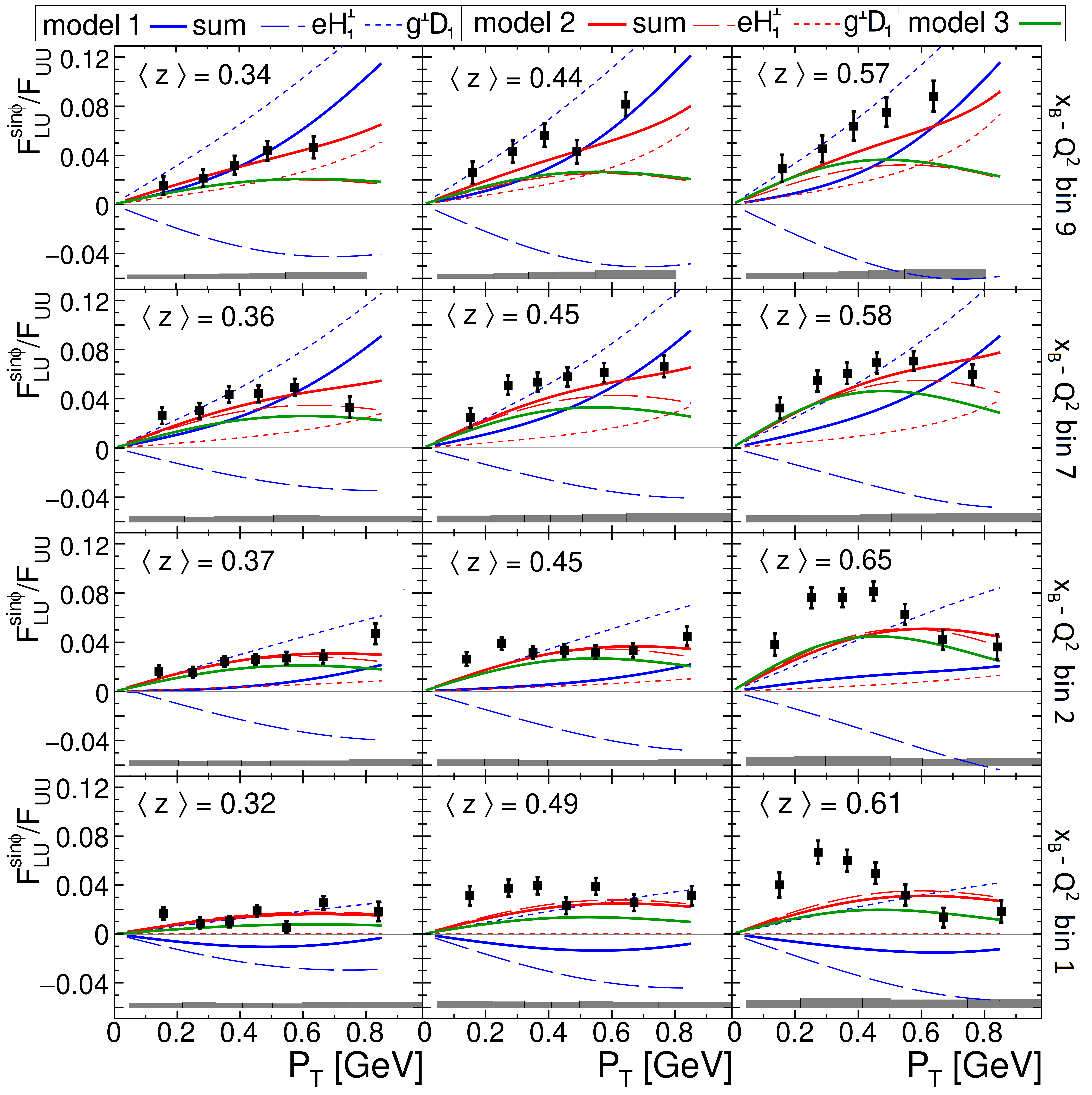}
	\caption{$P_{T}$ dependence of $F^{\sin\phi}_{LU}/F_{UU}$ for increasing $z$ bins (left to right) and for different $Q^{2}$-$x_{B}$ bins (see caption of Fig. \ref{fig:FLU_z_dep_multidim}). The systematic uncertainty is given by the grey histogram. The predictions of the different theoretical models are shown by the bold and dashed lines (see caption of Fig. \ref{fig:FLU_z_dep_multidim}).}
	\label{fig:FLU_pt_dep_multidim}
\end{figure}

The ratio $F^{\sin\phi}_{LU}/F_{UU}$ was extracted for each of the obtained 344 bins. The result for each bin, and the mean value of the kinematic variables in each bin, are listed in the supplemental material and in the CLAS physics database \cite{supl,CLAS_database}.
Figure\,\ref{fig:FLU_z_dep_multidim}  (\ref{fig:FLU_pt_dep_multidim}) shows the $z$ ($P_{T}$) dependence for selected $P_{T}$ ($z$) bins in different bins of $Q^{2}$, $x_{B}$, which represent the characteristics of the different kinematic regions. 
Evidently, the measured asymmetries are positive.
The experimental data reveal that the $z$ dependence changes from a more flat behavior at small $P_{T}$, $Q^{2}$, $x_{B}$ values to a steep increase at large $P_{T}$, $Q^{2}$, $x_{B}$.
Also for the $P_{T}$ dependence a small magnitude with a nearly flat behavior can be observed at small $Q^{2}$, $x_{B}$, $z$ values, while for increasing $z$ a peaking structure with varying mean value and width is observed at small  $Q^{2}$, $x_{B}$, while an increasing trend becomes dominant at large $Q^{2}$, $x_{B}$.
The results are compared to theoretical predictions, calculated using the models in Refs.\,\cite{MLEPJ13, MLEPJ14} (models 1 and 2) and Refs.\,\cite{PS03, kemal} (model 3).

The first two models describe the proton as an active quark plus (inert) spectator scalar and axial-vector diquarks. Both models include the $e H^{\bot}_{1}$ and $g^{\bot} D_{1}$ terms. 
The others are assumed to be small.
Model 1 uses a complicated propagator for the axial-vector diquark and a ratio for axial-vector/scalar strengths fitted  \cite{BCR08} to ZEUS \cite{Chekanov:2002pv} and GRSV01 \cite{Gluck:2000dy} PDFs.  Diquark masses and various cutoffs are also model parameters.  The PDF fits produce $|I=1,I_{z}=1\rangle$ and $|I=1,I_{z}=0\rangle$ axial-vector diquarks with very different masses. 
The model 1 diquark masses conflict with direct calculations \cite{Barabanov:2020jvn}, in which the scalar diquark mass is $\approx 80$\% of that associated with degenerate axial-vector correlations.  
Model 2 uses a simple propagator for axial-vector diquarks and the ratio of axial-vector and scalar is fixed by SU(4) spin-flavour symmetry. The model also uses mass degenerate axial-vector diquarks.
The models differ most significantly in the mass of the scalar diquark, the form of the propagator for the axial-vector diquark (complex versus simple) and the masses of these correlations (different versus degenerate).
The FFs used in both models are described in Ref. \cite{ZhunLu}.

Model 3 includes only the $e H^{\bot}_{1}$ term with the Collins function taken from the parametrization of Ref. \cite{Anselmino:2013vqa} and $e(x)$ based on a chiral quark soliton model \cite{PS03, OW04, Cebulla:2007ej}. In that model the $P_{T}$-dependence of the TMD $e(x, P_{T})$ is predicted to be narrow, emulating arguments elsewhere \cite{Schweitzer:2012hh}. Thus, in model-3 calculations it is assumed that $e(x,P_T) = e(x) \delta^{(2)}( P_T)$ \cite{kemal}. Hence, the approach explores the possibility that the FF's transverse momentum dependence is more important to the beam SSA than that of the target TMD.
This is the only model predicting the experimentally unmeasurable $\delta(x)$-contribution in $e(x)$ expected in QCD and related to the pion-nucleon sigma term \cite{Efremov:2002qh}.
Notably, since the Collins FF parametrizations in Refs.\,\cite{Anselmino:2013vqa, ZhunLu} are quite similar, then differences between this model's predictions and others stem largely from the form of $e(x)$ and the inclusion/omission of $g^\perp$.

%
The best reproduction of the general trend is provided by model 2. For this model, the comparison of the different kinematic regions shows that while $e H^{\bot}_{1}$ 
is dominant at small $Q^{2}$, $x_{B}$, quark-gluon correlations ($g^{\bot} D_{1}$) become increasingly important at large $Q^{2}$, $x_{B}$.
%
%
Model 1 mixes a complicated axial-vector diquark propagator with a simple proton wave function, uses a scalar diquark mass on the same scale as the lightest axial-vector diquark and vastly different masses for diquarks within the same isospin multiplet.  Here it is seen to be challenged by high precision experiments. In particular, its $e H_1^\perp$ contribution is uniformly negative.  If the scalar diquark mass is reduced to the value used in model 2, this conflict is eliminated. 
The simplicity and internal consistency of model 2 is a more natural beginning for phenomenology.
Even though model 3 uses a different approach, it provides results that are similar to the $e H^{\bot}_{1}$ term of model 2, which provides additional support for this model and, potentially, points to an important role for axial-vector diquarks in the proton's wave function \cite{Barabanov:2020jvn}.

In addition to the $z$ and $P_{T}$ dependence, a strong $x_{B}$ dependence (shown in the supplemental material) can be observed, with a more flat behaviour at small $z$, $P_{T}$ and an increasing trend for larger $P_{T}$, $z$ values. These kinematic dependencies can provide valuable insights into the kinematic dependence of the involved TMDs and FFs. 

Significantly, no parameters were varied in any of the models preparatory to making these experiment-theory comparisons, which therefore highlight the discriminating power of fully multidimensional analyses with high statistics over a wide kinematic range. Such data provide both the means of validating different models and their underlying assumptions, and the ability to place increasingly tight constraints on the TMDs involved.

Since a fully multidimensional analysis is herein made available for the first time, some new issues with model 2 are also exposed.  
(It was previously found \cite[Fig.\,6]{MLEPJ13} to underestimate the $\pi^0$ SSA data obtained by HERMES \cite{HERMES:2001hbj}.)
These things indicate that either the parametrizations of the involved TMDs and FFs have to be improved or that additional terms from Eq.\,\eqref{eqn:FLU_full} besides the two that have been used provide measurable contributions in some kinematic regions. Therefore, including the multidimensional data presented in this work will help to further constrain the TMDs and FFs in global fits.


In summary, the structure function ratio $F_{LU}^{\sin \phi}/F_{UU}$ corresponding to the polarized electron beam SSA in semi-inclusive deep inelastic scattering has been measured over a wide range of kinematics in a fully multidimensional study.
The comparison with calculations shows the promise of high-precision data to enable differentiation between competing reaction models and effects.  
%
In the context of currently available models, one sees:
the potential importance of the chiral odd distribution $e$, sensitive to emergent hadron mass, on the entire kinematic domain,
and a possible role for the poorly known T-odd chiral-even TMD $g^{\perp}$ at large $P_{T}$ and $z$;
and incipient new signals in support of a role for axial-vector diquark correlations in the proton’s wave function.
Therefore, including this multidimensional measurement into global fits, in combination with future measurements of unpolarized cross sections, as well as polarized target spin asymmetries, will provide new, strong constraints on the participating TMDs and FFs.
Such progress will set us firmly on the path to a deeper understanding of nucleon structure in the 3-D space most natural to picturing composite objects in relativistic quantum field theory.



We acknowledge the outstanding efforts of the staff of the Accelerator and the Physics Divisions at Jefferson Lab in making this experiment possible. We owe much gratitude to P. Schweitzer for many fruitful discussions concerning the interpretation of our results. This work was supported in part by the U.S. Department of Energy, the National Science Foundation (NSF), the Italian Istituto Nazionale di Fisica Nucleare (INFN), the French Centre National de la Recherche Scientifique (CNRS), the French Commissariat pour l$^{\prime}$Energie Atomique, the UK Science and Technology Facilities Council, the National Research Foundation (NRF) of Korea, the Helmholtz-Forschungsakademie Hessen für FAIR (HFHF), the Ministry of Science and Higher Education of the Russian Federation and the National Natural Science Foundation of China (grant no.\,12135007).  The Southeastern Universities Research Association (SURA) operates the Thomas Jefferson National Accelerator Facility for the U.S. Department of Energy under Contract No. DE-AC05-06OR23177.


\end{document}